\documentclass[preprint,showpacs,preprintnumbers,superscriptaddress,amsmath,amssymb]{revtex4}

\usepackage{graphicx}
\usepackage{dcolumn}
\usepackage{bm}

\begin{document}

\newcommand*{\cm}{cm$^{-1}$\,}
\newcommand*{\KFeSe}{K$_{0.8}$Fe$_{2-x}$Se$_2$\,}
\newcommand*{\Tc}{T$_c$\,}


\title{Nanoscale phase separation of antiferromagnetic order and
superconductivity in K$_{0.75}$Fe$_{1.75}$Se$_2$}

\author{R. H. Yuan}
\author{T. Dong}

\author{Y. J. Song}
\author{P. Zheng}

\author{G. F. Chen}

\affiliation{Beijing National Laboratory for Condensed Matter
Physics, Institute of Physics, Chinese Academy of Sciences,
Beijing 100190, China}

\author{J. P. Hu}

\affiliation{Department of Physics, Purdue University, West
Lafayette, Indiana 47907, USA}
\affiliation{Beijing National
Laboratory for Condensed Matter Physics, Institute of Physics,
Chinese Academy of Sciences, Beijing 100190, China}

\author{J. Q. Li}
\affiliation{Beijing National Laboratory for Condensed Matter
Physics, Institute of Physics, Chinese Academy of Sciences,
Beijing 100190, China}

\author{N. L. Wang\footnote{Correspondence and requests
for materials should be addressed to N.L.W.
(nlwang@aphy.iphy.ac.cn).}}
\affiliation{Beijing National
Laboratory for Condensed Matter Physics, Institute of Physics,
Chinese Academy of Sciences, Beijing 100190, China}

\begin{abstract}
We report an in-plane optical spectroscopy study on the
iron-selenide superconductor K$_{0.75}$Fe$_{1.75}$Se$_2$. The
measurement revealed the development of a sharp reflectance edge
below T$_c$ at frequency much smaller than the superconducting
energy gap on a relatively incoherent electronic background, a
phenomenon which was not seen in any other Fe-based
superconductors so far investigated. Furthermore, the feature
could be noticeably suppressed and shifted to lower frequency by a
moderate magnetic field. Our analysis indicates that this edge
structure arises from the development of a Josephson-coupling
plasmon in the superconducting condensate. Together with the
transmission electron microscopy analysis, our study yields
compelling evidence for the presence of nanoscale phase separation
between superconductivity and magnetism. The results also enable
us to understand various seemingly controversial experimental data
probed from different techniques.

\end{abstract}



\maketitle

The recent discovery of a new Fe-based superconducting system
A$_x$Fe$_{2-y}$Se$_2$ (A=alkaline metals or Tl, x$\leq$1) with
T$_c$ over 30 K has attracted considerable attention\cite{Guo}.
The system not only sets a new record for the highest T$_c$ for
the iron-selenide (FeSe-) based compounds at ambient pressure, but
also exhibits a number of physical properties that are markedly
different from all other Fe-pnictide/chalcogenide systems. Unlike
other Fe-pnictides or chalcogenides where the superconductors
develop from spin-density-wave (SDW) type metals with compensating
hole and electron Fermi surfaces locating respectively at the
Brillouin zone center and corners \cite{Dong,Cruz,Singh1}, the
superconductivity in this system was found to be in close
proximity to an insulating phase\cite{Fang}. The Fermi surface
(FS) topologies of superconducting compounds are very different
from previously known superconducting Fe-pnictides. Both band
structure calculations \cite{Nebrasov,Yan} and angle-resolved
photoemission spectroscopy (ARPES) studies \cite{Zhang1,Qian}
indicated that only the electron pockets are present in the
superconducting compounds, while the hole bands sink below the
Fermi level, indicating that the inter-pocket scattering between
the hole and electron pockets is not an essential ingredient for
superconductivity. More surprisingly, recent muon-spin relaxation
($\mu$SR) \cite{Shermadini,Pomjakushin}, neutron diffraction
\cite{Bao1,FYe,Bao2}, Raman\cite{AMZhang}, resistivity and
magnetization\cite{Liu} measurements on A$_x$Fe$_{2-y}$Se$_2$
(A=K, Rb, Cs, Tl) revealed a coexistence of superconductivity and
very strong antiferromagnetism. A blocked checkerboard
antiferromagnetic (AFM) order occurs in the Fe-deficient lattice
with the Fe vacancies forming a $\sqrt{5}\times\sqrt{5}\times1$
superstructure modulation. The ordered moment reaches 3.31
$\mu_B$/Fe \cite{Bao1,FYe,Bao2}.

A crucial question for the new system is whether the
superconductivity and the strong magnetism coexist on a
microscopic scale or they are phase separated? Neutron
diffractions indicated that the intensity of the magnetic Bragg
peaks shows a sharp downturn as the temperature is lowered below
T$_c$\cite{Bao1}. The Raman scattering experiments also revealed a
sudden reduction of the intensity of the two-magnon peak upon
entering the superconducting phase \cite{AMZhang}. Both
measurements seem to suggest that the antiferromagnetism and the
superconductivity are strongly coupled, pointing to a microscopic
coexistence of antiferromagnetic order and superconductivity. On
the other hand, several other measurements, including
magnetization \cite{BShen}, TEM \cite{ZWang}, ARPES \cite{FChen},
M\"{o}ssbauer \cite{Ryan}, STM \cite{WLi}, indicate that the
superconductivity and magnetic order are phase separated in the
sample. The strong inconsistency from different experiments
becomes a crucial issue for the system and needs to be solved.

Here we report optical spectroscopy measurements on
well-characterized superconducting samples. Unexpectedly, we
observed the development of a relatively sharp reflectance edge
below T$_c$ on the relatively incoherent electronic background.
Its energy scale is much smaller than the superconducting energy
gap, and as a consequence, this feature is not determined by the
pairing gap formation. Furthermore, the feature could be
noticeably suppressed and shifted to lower frequency by a moderate
magnetic field. We elaborate that this reflectance edge arises
from the development of a Josephson-coupling plasmon in the
superconducting condensate. The data highly suggest a nanoscaled
and possibly stripe-type phase separation between
superconductivity and magnetic insulator, which was further
confirmed by the transmission electron microscopy (TEM) dark-field
image technique. The results also enable us to understand various
seemingly controversial experimental data probed from different
techniques.

\section{Results}

Figure 1 shows the temperature dependence of the in-plane
resistivity, magnetization, and specific heat data. The
resistivity shows a weak metallic temperature dependence. Two step
transitions were seen in resistivity curve. A sharp drop at 42 K
was observed followed by a major transition near 30 K. The
transition at 42 K could be weakly seen in a highly enlarged scale
in the magnetization curve with H$\parallel$c, similar to the
report in Ref. \cite{DMWang}, but is not visible in the specific
heat measurement, suggesting an extremely small fraction or
interface superconductivity at this transition temperature in the
sample. Clear diamagnetization in susceptibility and jump in
specific heat were seen at lower temperature, $\sim$28 K, where
the sample already reaches zero resistivity. High temperature
magnetization measurement revealed the presence of an AFM phase
transition near 520 K for this sample.

Figure 2 shows the R($\omega$) and $\sigma_1(\omega)$ spectra for
the K$_{0.75}$Fe$_{1.75}$Se$_2$ sample. The left panels show the
R($\omega$) and $\sigma_1(\omega)$ spectra up to 8000 \cm, the
right panels show the spectra in the expanded low frequency region
within 250 \cm. Similar to the insulating compound with lower Fe
content \cite{Chen1}, the reflectance over broad frequencies is
rather low, roughly below the value of 0.4. In the earlier study
on the insulating compounds, two characteristic spectral features
specific to the K$_x$Fe$_{2-y}$Se$_2$ system were identified: a
double peak absorption structure between 4000-6000 cm$^{-1}$ and
abundant phonon peaks (much more than those expected for a
standard 122 structure). Both features were interpreted to be
highly related to the blocked checkerboard AFM order associated
with the presence of Fe vacancies and their orderings
\cite{Chen1}. Those features are also seen in the present
compound, suggesting the presence of Fe vacancies and their
orderings in the superconducting samples. The presence of
$\sqrt{5}\times\sqrt{5}\times1$ superstructure modulation was
confirmed by the TEM measurement as we shall present below. It is
also consistent with the high temperature magnetization
measurement showing the presence of AFM transition near 520 K
(Fig. 1 (d)).

The major spectral change relative to the insulating compound
appears at low frequencies. The reflectance R($\omega$) values at
low frequencies are obviously higher than that of insulating
samples \cite{Chen1}. Furthermore, the low-$\omega$ R($\omega$)
shows a metallic temperature dependence: the R($\omega$) values
increases with decreasing temperature. However, this kind of
metallic response is rather weak. In the optical conductivity
spectra, the low-frequency region is still dominated by phonon
modes. The electronic background has a rather low spectral weight
without showing a clear Drude-like component. Unexpectedly, a
relatively sharp reflectance edge develops below 30 \cm in the
superconducting state. This surprising sharp feature was
repeatedly observed in different superconducting samples.

The observation of a sharp reflectance edge below T$_c$ is the
most intriguing experimental result in the infrared spectroscopy
measurement. It is important to understand its physical origin.
Naturally, one has to examine whether or not the spectral feature
is caused by the formation of a superconducting energy gap? As
indicated below, this possibility is highly unlikely for several
reasons. First and the most importantly, the superconducting
energy gap (2$\Delta$) amplitudes determined directly by the ARPES
experiments on the electronic pockets, which are the only
dispersive bands crossing the Fermi level, are close to 18-20 meV
\cite{Zhang1,LZhao}. Those values are much larger than the energy
scale seen for the edge. Second, although the sample is
superconducting, the reflectance values at the lowest measurement
frequency limit in the normal state are still far below the unit,
leading to a non-Drude-like response in $\sigma_1$($\omega$). As
we shall also explain below that the sample likely contains some
insulating phase, it is hard to imagine that a full gap feature
could be realized in the relatively nonhomogeneous sample.
Furthermore, the reflectance spectral at 8 K shows a strong dip
feature near 44 \cm. At this frequency, its R($\omega$) value is
much lower than that in the normal state (R(8 K)/R(35
K)$\approx$0.9). The dip is so pronounced that the feature is
unlikely to be related to a superconducting gap.

On the other hand, the relatively sharp feature is more likely to
be caused by the Josephson-coupling plasma edge. As seen from Fig.
3, the real part of the dielectric function $\epsilon_1$($\omega$)
in the normal state (e.g. at 35 K) is positive and further
increases in the lowest-frequency region, similar to the case of
insulating dielectrics. However, in the superconducting state, the
low-$\omega$ $\epsilon_1$($\omega$) becomes a rapidly decreasing
function of $\omega$ toward $\omega$=0. The reflectance edge
corresponds to the zero-crossing of the real part of the
dielectric function $\epsilon_1$($\omega$), indicating that this
edge is resulted from collective plasma oscillation.

The Josephson plasmon has been widely observed in the high-T$_c$
cuprates with the electric field polarized along the c-axis
\cite{Tamasaku,Basov,Uchida,Shibata}. Since the cuprate
superconductor could be viewed as an alternating stack of
superconducting CuO$_2$ planes and insulating building blocks, the
optical response is still insulator-like in the normal state.
However, once entering into the superconducting state, those
CuO$_2$ layers are coupled through the Josephson tunneling effect.
Then, a plasma edge corresponding to the superconducting
condensate emerges with its location mainly determined by the
critical current density in the stacking direction. Because of the
periodic spacing of the CuO$_2$ layers in the crystal structure,
the Josephson plasma edge is very sharp. We suggest that similar
situation occurs for K$_{0.75}$Fe$_{1.75}$Se$_2$ compound here. If
the sample has a nanoscale phase separation between the
superconducting and insulating phases, for example, a stripe-type
phase separation as outlined schematically in the inset of Fig. 3,
this type of Josephson coupling plasmon would be well expected.

To verify the above proposal, we performed TEM investigations by
using the superstructure reflection spots for dark-field imaging
on the superconducting crystals. Figures 4 (a) shows an electron
diffraction pattern taken along the [001] zone axis direction, in
which the superstructure spots with both (1/5, 3/5, 0) and (1/2,
1/2, 0) wave vectors can be clearly seen. Figure 4 (b) displays
the dark-field TEM images by using one of the superstructure spot
as indicated in Fig. 4 (a). The well ordered regions with fine
striped or speckled contrasts can be commonly observed. The length
of a stripe could be in the range from several to several tens or
even over 100 nanometers, while it width is usually less than 10
nanometers. This kind of complex contrast in Fig. 4 (b) can be
explained directly as the coexistence of Fe-vacancy ordered bright
areas and Fe deficiency-free (or Fe-disordered) areas. Taking into
account of the remarkable superconductivity in present sample, we
can conclude that the Fe deficiency-free/Fe-disordered areas are
mainly governed by the superconducting phase. The dark-field TEM
image technique provides direct and strong support for the
presence of nanoscaled stripe or speckled phase separation.
Because the phase separation could not be uniform everywhere in
the sample, the Josephson coupling strength would exhibit local
variations. It implies that the Josephson plasma frequencies would
show a distribution around a center frequency. As a result, the
Josephson plasma edge could not be as sharp as that observed in
the c-axis optical response in the cuprates.

Naturally, the insulating phase could be assigned to the AFM
ordered phase with a $\sqrt{5}\times\sqrt{5}\times1$
superstructure modulation, while the superconducting phase
originates from the K-deficient K$_{0.75}$Fe$_2$Se$_2$
composition. Because of the K-vacancy ordering, it results in a
$\sqrt{2}\times\sqrt{2}\times1$ superlattice. This phase is indeed
heavily electron-doped, and thus has a big electron FS, which has
been detected by the ARPES experiment \cite{Zhang1,Qian}. From
earlier optical measurement on the insulating sample, a small
indirect gap $\sim$30 meV was identified \cite{Chen1}. Because the
barrier is rather low, the two superconducting stripes separated
by the AFM ordered insulating region could be coupled through the
Josephson tunnelling effect in the superconducting state.

To further substantiate the picture, we tried to reproduce the
shape of the Josephson plasma edge with a simple model as
suggested by van der Marel and Tsvetkov \cite{vanderMarel} that
has taken account of the distribution of the Josephson plasma
frequencies around a center frequency. The expression for
dielectric function has the form of
\begin{equation}
\label{Dielectricfunc}
  \frac{1}{\varepsilon_J(\omega)} = \int
  dX\frac{F(X)\omega^2}{\varepsilon_\infty(\omega^2-X^2)+4\pi
  i\omega\sigma_n},
\end{equation}
where $F(X)$ is the normalized distribution function of the
screened Josephson plasma frequencies, which we assume to have a
form of Gaussian distribution function, $F(X)$ =
$1/(\sigma\sqrt{2\pi})exp(-(X-\omega_J)^2/2\sigma^2)$. In the
above equation, $\omega_J$ is the central frequency of the
screened Josephson plasma frequency, $\sigma_n$ the normal fluid
component, $\varepsilon_\infty$ the high frequency dielectric
constant, and $\sigma$ the variance of the distribution function.
Figure 5 (a) shows the shape of the Josephson plasma edge as a
function of the variance of the distribution function in the
normalized Gaussian distribution function of the screened plasma
frequencies. The change of the normalized Gaussian distribution of
the screened plasma frequencies with the parameter $\sigma$ is
plotted in Fig. 5 (b). Here the parameters were chosen as
$\omega_J$=40 \cm, $\sigma_n$=5 S/cm, $\varepsilon_\infty$=15. The
variance of the distribution function is chosen as $\sigma$=1, 5,
10, 15. Clearly, the Josephson plasma edge becomes less sharp when
the Josephson plasma frequencies have broader distributions.
Qualitatively, it explains the observation fairly well.

Since the Josephson coupling plasmon is a phenomenon related to
the tunnelling of the condensed superconducting carriers, it
should be easily influenced by the external magnetic field. We
therefore explored the effect of the magnetic field on the
Josephson coupling plasmon edge. Figure 6 shows a far-infrared
reflectance measurements under zero field and H=8 T for a
different sample grown in the same condition. The field is applied
along the c-axis which is perpendicular to the electric field of
the infrared radiation. Relative to the curve at 35 K, the
R($\omega$) at 5 K shows a clear edge-like shape with a dip
appearing near 50 \cm. Applying magnetic field to the sample, the
edge-like feature weakens and shifts towards lower frequencies.
From Fig. 6 (a), it is easy to find that R($\omega$) under the
field of 8 T at T=5 K follows the normal-state R($\omega$)
measured at 35 K down to much lower frequencies. Figure 6 (b) is a
plot of the ratio of the zero-field reflectance at 35 K to the
reflectance curves at 5 K under different fields. Then we find a
peak in the ratio curve, which shifts to lower frequency side by
over 13 \cm by a magnetic field of 8 T. Meanwhile, the intensity
of the peak drops. The rather significant change of the plasma
edge structure by such a moderate field also favors a Josephson
coupling plasmon scenario rather than a superconducting energy
gap, since the upper critical field is known to be extremely high
in this compound \cite{DMWang}.

\section{Discussion}

Our experimental investigation revealed novel Josephson coupling
phenomenon in the new A$_x$Fe$_{2-y}$Se$_2$ superconducting single
crystals, which was not seen in other Fe-based superconductors. In
fact, except for the polarized infrared measurement of
\textbf{E}$\parallel$c-axis on cuprates below T$_c$, we were not
aware of observation of Josephson plasmons in any other
superconducting compounds. Our experimental results yield
compelling evidence for the presence of nanoscale phase separation
between superconducting and AFM ordered insulating phases. The
rather incoherent low-frequency optical conductivity is naturally
due to the presence of sizeable fraction of insulating phase which
largely blocks the conducting paths. The presence of the sizeable
fraction of insulating phase could also account for the presence
of double interband transition peaks between 4000 and 6000 \cm and
abundant phonon peaks observed in all measured superconducting
samples, which were interpreted as being associated with the
blocked antiferromagnetism due to the presence of Fe vacancy
ordering \cite{Chen1}.

Based on the nanoscale phase separation picture, one can also
explain those seemingly controversial experimental data probed
from different techniques. Currently, the strongest experimental
support for a microscopic coexistence of antiferromagnetic order
and superconductivity comes from the neutron diffraction
\cite{Bao1} and two-magnon Raman-scattering \cite{AMZhang}
measurements. The intensity of the magnetic Bragg peaks in neutron
diffraction shows a sharp downturn (approximately 5$\%$) as the
temperature is lowered below T$_c$. The intensity of the
two-magnon peak in Raman scattering also undergoes a 5$\%$ sudden
reduction on entering the superconducting phase. If the phase
separation occured at a macroscopic region, this drop is really
hard to understand. However, the phase separation revealed in this
study comes out at a nanoscale level. There exist a huge amount of
the phase boundaries. Below T$_c$, the superconducting proximity
effect near the phase boundaries would effectively reduce the AFM
ordered region. Then, the reduction of several percent magnetic
response would not be unexpected.

\section{Methods}
The single crystals used in the present study were grown from a
self-melting method with nominal concentration of
K:Fe:Se=0.8:2.1:2 in a procedure similar to the description in
reference \cite{Chen1}. The actual composition, determined by the
average value of the energy dispersive x-ray (EDX) spectroscopy
analysis, was found very close to 0.75:1.75:2. The temperature
dependence of the in-plane resistivity, magnetization, and
specific heat were measured by PPMS and SQUID from Quantum Design.
Optical measurements at zero and magnetic field were done on
Bruker 113v and Vertex 80v spectrometers in the frequency range
from 17 to 25000 cm$^{-1}$. The sample surface area is about
4mm$\times$4mm. An \textit{in situ} gold and aluminum overcoating
technique was used to get the reflectance \emph{R}($\omega$). The
real part of conductivity $\sigma_1(\omega)$ is obtained by the
Kramers-Kronig transformation of \emph{R}($\omega$). The TEM
measurement was taken on a FEI Tecnai-F20 (200 kV) transmission
electron microscope.

\begin{acknowledgments}
We acknowledge enlightening discussions with Christian Bernhard.
This work was supported by the NSFC and the 973 project of the
MOST.
\end{acknowledgments}

\subsection{Author contributions} N.L.W. planned and coordinated the experiments.
R.H.Y, T.D. grew single crystals and carried out resistivity,
magnetization, specific heat, and optical spectroscopy
experiments. T.D. did the model simulation. Y.J.S. J.Q.L.
performed TEM measurement and dark field analysis. P.Z. helped
with transport and specific heat measurement. G.F.C. contributed
to crystal growth at the early stage. J.P.H. helped with data
interpretation. N.L.W. wrote the paper.

\subsection{Additional Information} The authors declare no
competing financial interests.


\begin{figure}[t]
\includegraphics[width=9cm]{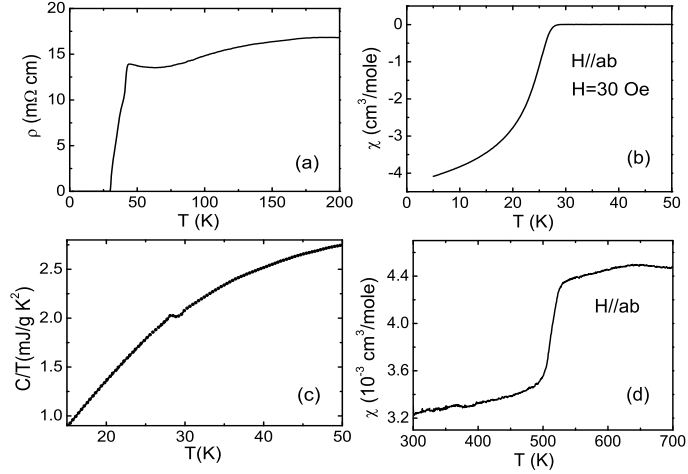}
\caption{(a) The in-plane resistivity versus temperature for
K$_{0.75}$Fe$_{1.75}$Se$_2$ single crystal. The sample shows a
sharp drop in $\rho$(T) at 43 K, then a major transition near 30
K. (b) The temperature dependence of magnetic susceptibility below
50 K. Sharp diamagnetic transition appears at 28 K. (c) The low
temperature specific heat data for the sample. Clear specific jump
is observed near 28 K, evidencing bulk superconductivity. (d) The
high temperature magnetic susceptibility measured at 1 T for the
sample. Antiferromagnetic phase transition is still present for
the superconducting sample.}
\end{figure}

\begin{figure}
\includegraphics[width=9cm]{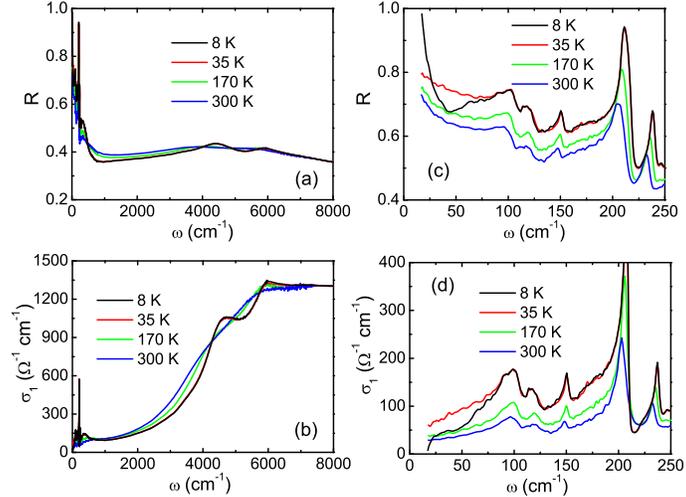}
\caption{(Color online) (a) and (b): optical reflectance
R($\omega$) and conductivity $\sigma_1(\omega)$ spectra at
different temperatures up to 8000 \cm. (c) and (d): An expanded
plot of R($\omega$) and $\sigma_1(\omega)$ spectra below 250 \cm.
A sharp plasma edge in R($\omega$) develops at low frequency in
the superconducting state.}
\end{figure}

\begin{figure}
\includegraphics[clip,width=6.5cm]{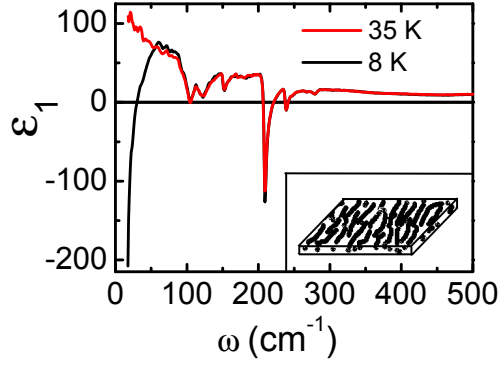}
\caption{(Color online) The real part of dielectric functions
versus frequency at 35 and 8 K. $\epsilon_1$($\omega$) at 35 K is
positive and increases further with decreasing frequency. However,
the $\epsilon_1$($\omega$) crosses the zero at low frequency in
the superconducting state. The inset shows a schematic picture of
the nanoscale stripe-type phase separation between superconducting
(white stripe) and insulating (black) regions.}
\end{figure}

\begin{figure}[t]
\includegraphics[clip,width=6.5cm]{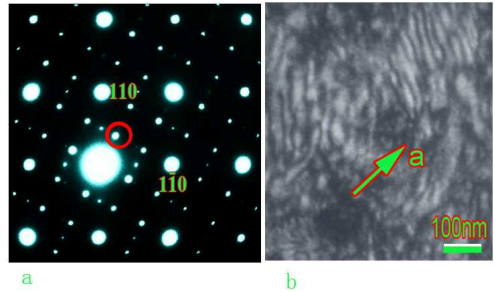}
\caption{(Color online) Phase separation in a
K$_{0.75}$Fe$_{1.75}$Se$_2$ superconducting crystal. (a) Electron
diffraction pattern showing the presence of superstructure spots
along the [310] direction, the cycled spot is used for dark-field
imaging. (b) Dark field image taken from a thin region of a
K$_{0.75}$Fe$_{1.75}$Se$_2$ crystal. Stripe-type phase separation
could be directly observed. The arrow \textbf{a} indicates the
[100] direction.}
\end{figure}

\begin{figure}
\includegraphics[clip,width=11cm]{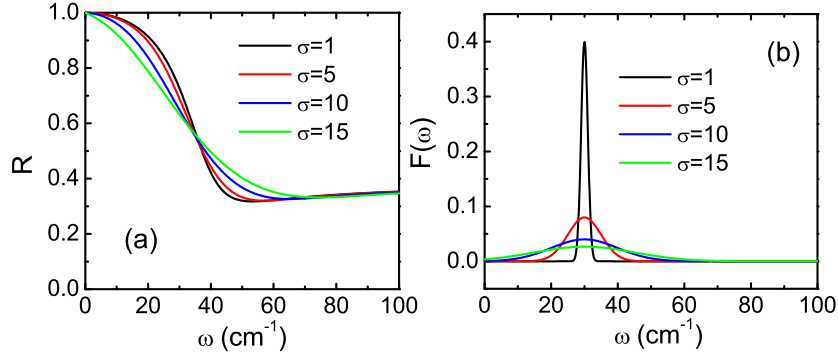}
\caption{(Color online) (a) the variation of the reflectance edge
shape as a function of the distribution of the the Josephson
plasma frequencies. (b) the plot of the normalized Gaussian
distribution of the screened plasma frequencies as a function of
the parameter $\sigma$.}
\end{figure}

\begin{figure}
\includegraphics[clip,width=11cm]{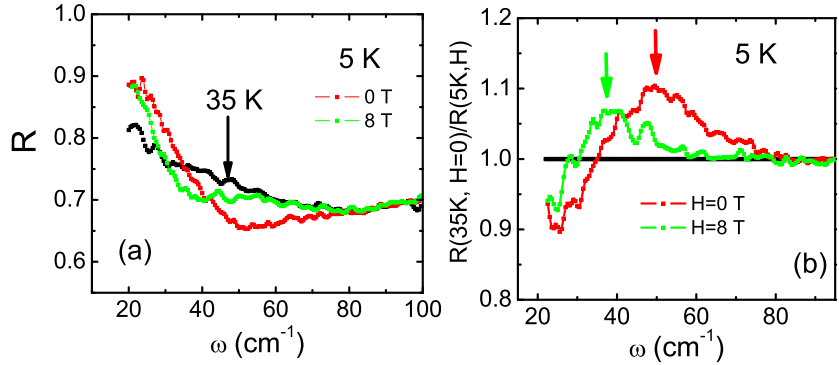}
\caption{(Color online) (a) The reflectance at 5 K under H=0 and 8
T. The reflectance curve at 35 K under zero field is also shown as
indicated by the black arrow. (b) The ratio of the zero-field
reflectance at 35 K to the reflectance curves at 5 K under
different fields. The arrows indicate the frequency shift of the
peaks.}
\end{figure}


\end{document}